\documentclass[conference]{IEEEtran}\usepackage[]{graphicx}\usepackage[]{color}
\makeatletter
\def\maxwidth{ %
  \ifdim\Gin@nat@width>\linewidth
    \linewidth
  \else
    \Gin@nat@width
  \fi
}
\makeatother

\definecolor{fgcolor}{rgb}{0.345, 0.345, 0.345}

\usepackage{framed}
\makeatletter
 {\par\unskip\endMakeFramed%
 \at@end@of@kframe}
\makeatother

\definecolor{shadecolor}{rgb}{.97, .97, .97}
\definecolor{messagecolor}{rgb}{0, 0, 0}
\definecolor{warningcolor}{rgb}{1, 0, 1}
\definecolor{errorcolor}{rgb}{1, 0, 0}
\newenvironment{knitrout}{}{} 

\usepackage{alltt}\usepackage[]{graphicx}\usepackage[]{color}
\makeatletter
\def\maxwidth{ %
  \ifdim\Gin@nat@width>\linewidth
    \linewidth
  \else
    \Gin@nat@width
  \fi
}
\makeatother

\usepackage{alltt}
\usepackage{cite}

  \usepackage{graphicx}

\usepackage{amsmath}
\usepackage{algorithmic}
\usepackage{textcomp}

\usepackage{gensymb}

\usepackage{balance}

\usepackage{dblfloatfix}  

\usepackage[nolist]{acronym}

\IfFileExists{upquote.sty}{\usepackage{upquote}}{}
\IfFileExists{upquote.sty}{\usepackage{upquote}}{}
\usepackage[%
	IEEE,%
	]{copyright-overlay}

\CopyrightAuthor{Christian Arendt}
\CopyrightYear{2021}
\ConferenceName{2021 IEEE 94th Vehicular Technology Conference (VTC-Fall)}
\Doi{}
\Bibtex{%
	@InProceedings\{Arendt2021pushing,\textCR\textLF
	\textHT{}Author = \{Christian Arendt and Manuel Patchou and Stefan Böcker and Janis Tiemann and Christian Wietfeld\},\textCR\textLF
	\textHT{}Title = \{Pushing the Limits: Resilience Testing for Mission-Critical Machine-Type Communication\},\textCR\textLF
	\textHT{}Booktitle = \{2021 IEEE 94th Vehicular Technology Conference (VTC-Fall)\},\textCR\textLF
	\textHT{}Publishingstatus = \{accepted for presentation\},\textCR\textLF
	\textHT{}Address = \{Virtual Event\},\textCR\textLF
	\textHT{}Month = \{sep\},\textCR\textLF
	\}
}

\begin{document}

\begin{acronym}
\acro{5G}[5G]{5th Generation of mobile networks}
\acro{STING}[STING]{Spatially Distributed Traffic and Interference Generation}
\acro{MQTT}[MQTT]{Message Queuing Telemetry Transport}
\acro{SUT}[SUT]{System Under Test}
\acro{AP}[AP]{Access Point}
\acro{FPV}[FPV]{First Person View}
\acro{IAT}[IAT]{Inter Arrival Time}
\acro{ASTM}[ASTM]{American Society for Testing and Materials}
\acro{LOS}[LOS]{Line of Sight}
\acro{NLOS}[NLOS]{Non Line of Sight}
\acro{MJPEG}{Motion JPEG}
\acro{UGV}{Unmanned Ground Vehicle}
\acro{ROS}{Robot Operating System}
\acro{LTE}{Long Term Evolution}
\acro{MTC}{Machine-Type Communication}
\acro{URLLC}{Ultra-Reliable Low Latency Communication}
\acro{DUT}{Device under Test}
\acro{DUTs}{Devices under Test}
\acro{Multi-RAT}{Multiple Radio Access Technology}
\end{acronym}

%
 \title{Pushing the Limits: Resilience Testing for\\Mission-Critical Machine-Type Communication}

%
\author{\IEEEauthorblockN{Christian Arendt, Manuel Patchou, Stefan B\"ocker, Janis Tiemann, Christian Wietfeld}
	\IEEEauthorblockA{TU Dortmund University, Germany\\
		Communication Networks Institute (CNI)\\
		Email: \{christian.arendt, manuel.mbankeu, stefan.boecker, janis.tiemann, christian.wietfeld\}@tu-dortmund.de}}


%


\maketitle

\PrintCopyrightOverlay
\begin{abstract}
  
Interdisciplinary application fields, such as automotive, industrial applications or field robotics show an increasing need for reliable and resilient wireless communication even under high load conditions.
These mission-critical applications require dependable service quality characteristics in terms of latency and especially stability. Current deployments often use either wired links that lack the flexibility to accommodate them, or wireless technologies that are susceptible to interference. Depending on the application and the surrounding environments, different technologies can meet the associated requirements and have to be tested deliberately to prevent unexpected system failure. \\
To stress test these infrastructures in a reproducible and application-aware manner, we propose STING, a spatially distributed traffic and interference generation framework. 
STING is evaluated in a remote control test case of an Unmanned Ground Vehicle that serves as a scout in Search and Rescue missions. A significant impact of interference on the remote control quality of experience is shown in tests with different operators, which result in an 80\% increase in completion time in our test scenario with high interference on the radio channel. With this case study, we have proven STING to be a reliable and reproducible way to asses resilience against interference of wireless machine-type communication use cases. Our concept can find use for any type of wireless technology, in unlicensed (e.g. Wi-Fi) as well as licensed bands (e.g. 5G).
\end{abstract}


%
\IEEEpeerreviewmaketitle

\section{Introduction}
\label{sec:intro}
\begin{figure}[t]

	\centering
	\includegraphics[width = .49\textwidth, trim = 0mm 0mm 0mm 0mm, clip]
	{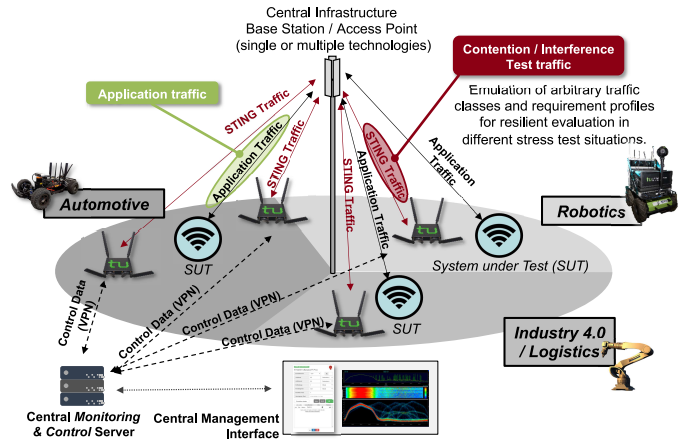}
	\caption{Spatially Distributed Traffic and Interference Generation (STING): Cross-domain application of STING concept for mission-critical machine-type communication resilience testing
}

	\label{fig:intra}
\end{figure}

Wireless connectivity is becoming a more and more crucial aspect of various applications in the industrial domain as well as search and rescue applications. Especially in the field of robotics and (semi-)autonomous systems, the network performance needed to communicate sensor and control data near real time has to be reliably available independent from the surrounding environment in order to enable mission-critical applications and pose a real benefit for vehicle operators and industries alike. Opposed to the ongoing rollout of the \ac{5G}, which is said to fulfill many of the needs of today's \ac{MTC} applications, current implementations often rely on Wi-Fi for its ease of use and cost-efficient deployment. However, especially in the aforementioned domains, reliability and guaranteed service quality are critical, and while every wireless communication is in principle prone to interference by other (known or unknown) transmissions in the same frequency range, this is especially true for those in unlicensed frequency bands like Wi-Fi. Future developments incorporate application-aware connectivity solutions and even \ac{Multi-RAT} approaches \cite{mahmood20206g}, which will have to prove their suitability for specific mission-critical \ac{MTC} applications in high load scenarios. Therefore, deliberate resilience testing and evaluation depending on the targeted application is necessary to avoid malfunctioning systems in advance. This evaluation is often carried out through simulations, which naturally do not take every aspect of a deployment situation into account.  
This work therefore proposes \ac{STING}, a modular framework for stress testing network infrastructures with distributed devices as shown in Fig. \ref{fig:intra}, generating configurable traffic patterns which can emulate real-world applications.   
\ac{STING} consists of a central management entity and distributed devices, which can emulate and generate various real-world traffic patterns to enable different kinds of stress test scenarios.    

The present work is structured as follows: Section \ref{sec:rel_work} gives an overview of related and previous works. Section \ref{sec:architecture} describes the proposed system architecture and its capabilities for stress testing network infrastructures. Section \ref{sec:application} shows a case study of a proposed rescue robotics test case scenario in which the influence of Wi-Fi interference on the operability of a teleoperating robotic system is analyzed. Results of this analysis are shown in section \ref{sec:contention}. Section \ref{sec:concl} finally concludes this work.   

\section{Related Work}
\label{sec:rel_work}
Mission-critical \ac{MTC} has been a major driver for development of 5G technologies for \ac{URLLC} service classes, as its high requirements pose a hard challenge for wireless deployments \cite{MohammedMissionCriticalMachineTypeCommunication2019}. The evolution of mission-critical \ac{MTC} towards 6G has been addressed in \cite{mahmood20206g}, which discusses future use cases as well as solutions to meet the ever-growing demands of such applications. We propose the \ac{STING} framework as a measure to reliably test these upcoming infrastructures for resilience even in high load scenarios. \\
The concept of this work is based on \cite{kaulbarsSpatiallyDistributedTraffic2015}. The authors implemented a distributed traffic generation system for stress-testing networks at locations of smart meters and evaluated the system utilizing LTE. 
The authors of \cite{AlvarezMulticlientemulationplatform2016} used a similar approach to assess the performance of schools IEEE\,802.11n based Wi-Fi networks prior to the distribution of laptops to students. They also used iPerf as a tool for performance measurements; however, the emulated devices are strictly colocated as the emulation occurs in a central unit.
A generalized approach to interference testing is discussed in \cite{KoepkeInterferenceCoexistenceWireless2015}. The authors focus on reproducible tests in a controlled environment using single \ac{DUTs}, whereas this work focuses on application-oriented test environments. 

Performance Testing of robot systems with regard to communication has been addressed in \cite{PatchouHardwareSimulationLoop2020} following a hardware-in-the-simulation-loop approach. Apart from the abstraction natural to a simulation, the work focused on network coverage, where interference was not taken into account. \\
%
%
The importance of network communications within robot missions has been shown in works such as \cite{gueldenring2020} where unmanned aircraft systems were used beyond \ac{LOS} to carry out search and rescue missions while relying on a robust and long-range suitable multi-link, which resulted from the aggregation of a self-deployed \ac{LTE} cell and the services of two mobile network operators. A compilation of robotic application examples having a critical reliance on network communications is given \cite{Queralta2020} with the focus set on search and rescue scenarios featuring the collaborative multi-robot teams. The various challenges and constraints that different robot types (e.g. ground, aerial, surface, or underwater) encounter in different environments (e.g. maritime or post-disaster scenario) are also discussed. This stresses the relevance of the testing and certification processes of these robotic systems.\\
%
%
In \cite{Majzik2019}, the system level testing of autonomous vehicles is addressed to cover the system behavior in complex traffic scenarios, which may be left unaccounted for by component-level testing. This stresses the needs for application-aware testing based on real-world scenarios in addition to isolated tests of specific components.   
%
%
Critical operations such as emergency responses and rescue missions require that the deployed robots possess certain sets of capabilities, thus stressing the importance of standards.
To address this, NIST researchers are collaborating with others to establish a collection of test suites under the standards development organization ASTM International which can be used to challenge specific robot capabilities in repeatable ways in order to facilitate direct comparison of different robots as well as particular configurations of similar robot models\cite{Jacoff2010}.\\
While these various test methods for different aspects and capabilities of robotic systems exist, to our knowledge there are no standardized test methods specific to communication performance under high load situations.\\

\section{Spatially Distributed Traffic and Interference Generation: Architecture}
\label{sec:architecture}
\begin{figure}[htb]
	\centering
	\includegraphics[width = .48\textwidth, trim = 0mm 0mm 0mm 0mm, clip]
	{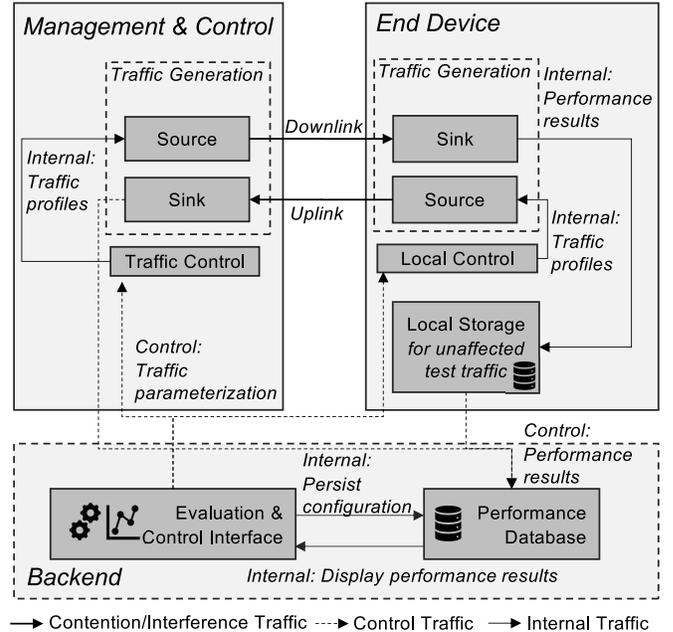}
	\caption{Overview of \ac{STING} system architecture}
	\label{fig:architecture}
\end{figure}

The \ac{STING} system is composed of two main parts as depicted in Fig. \ref{fig:architecture}, a central management and control system and the distributed end devices used for traffic generation, which will be briefly described in the following sections. These entities can be used with different and multiple network infrastructures.

\subsection{Management and Control Server and backend}
The system's central server acts as a counterpart to the end devices for traffic generation. The system is configured using a web interface on the central server in which traffic parameters are defined. The configurations are then published to a topic on an \ac{MQTT} broker running on the server to which the end devices are subscribed (cf. Fig. \ref{fig:architecture}). Performance metrics about the transmissions are then published back to the server and persisted in a database for further analysis. With this, local or global shortcomings of a network infrastructure can be determined and fixed before an actual application is deployed. 

\subsection{STING End Devices}
The end devices are based on an embedded PC specialized in network applications, allowing the usage of various network adapters in the form of M.2 module cards. In the case study discussed later in this work, an Intel AX200 Wi-Fi 6 module is used \cite{IntelIntelWiFiAX200}. 
The devices can be deployed in a distributed test scenario, e.g. a manufacturing hall, and emulate traffic occurrence of one or multiple target applications. Performance metrics such as throughput, latency, or round trip time of the established traffic flow are stored locally so that the results transmissions do not distort the tests. After a test scenario is finished, results are published to the central server using \ac{MQTT} to be persisted.

\subsection{Traffic Generation Concept}
\begin{figure}[htb]
	\centering
	\includegraphics[width = .48\textwidth, trim = 0mm 0mm 0mm 0mm, clip]
	{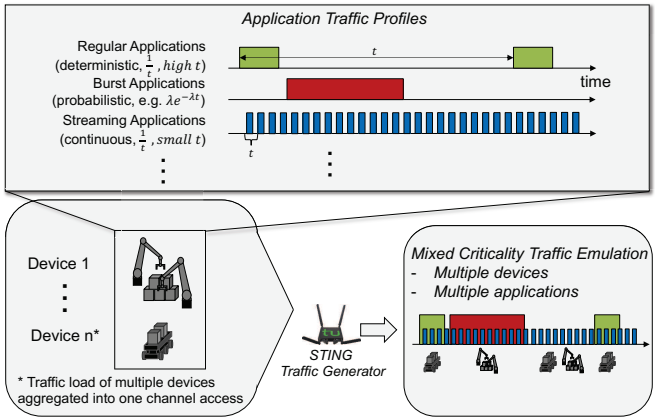}
	\caption{Traffic Generation concept based on aggregation of application traffic demands}
	\label{fig:trafficmodel}
\end{figure}

The \ac{STING} platform can be used to generate various heterogeneous network traffic constellations. This enables reproducible stress-tests based on constant streaming interference as discussed later in section \ref{sec:application} as well as scalability tests based on the emulation of real application traffic of the \ac{SUT}. Therefore one end device can emulate multiple applications and traffic patterns, as depicted in Fig. \ref{fig:trafficmodel}. 
Applications are emulated in separate processes. Therefore, regular small data packets such as sensor data and bursty file transmissions are generated in a per-packet way following an underlying pattern or probability distribution function. In contrast, constant streaming applications like video data can be emulated via the open-source traffic generator iPerf \cite{iPerfultimatespeed}. 
These application types are aggregated and sent from the end devices to the management and control system or vice versa, resulting in either uplink or downlink traffic depending on the application.
This approach enables analysis of various application types and requirements with different demands regarding latency and reliability, leading to an emulation of mixed-criticality applications \cite{mahmood20206g}. 

\section{Case Study: Robotic Test Case}
\label{sec:application}
An implementation of the \ac{STING} system is used to build an interference stress-test scenario for teleoperated rescue robotic systems. Robotic systems promise a high value for search and rescue applications in hazardous environments. However, this can only be achieved with a reliable wireless communication link. Due to their cost-effective and easy-to-use deployment, Wi-Fi systems are an attractive solution for this type of application. However, in order to validate the performance of these systems beforehand, stress-test methods are crucial. Standard test methods as provided for example by the \ac{ASTM}, only cover test methods for 1-to-1 \ac{LOS} \cite{ASTME285412Standard} and \ac{NLOS} \cite{ASTME285512Standard} scenarios. Therefore, we propose a test case based on the \ac{STING} system. 

\subsection{Test Case Scenario}
\begin{figure}[htb]
	\centering
	\includegraphics[width = .48\textwidth, trim = 0mm 0mm 0mm 0mm, clip]
	{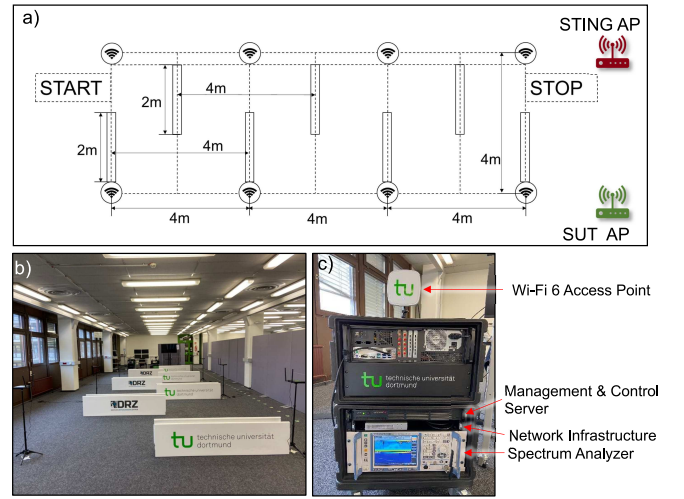}
	\caption{STING implementation for robotic test case study. a) Schematic overview of the robotic parcours. b) Real implementation. c) STING Management \& Control Unit}
	\label{fig:testcase}
\end{figure}

The test field used for this purpose is shown in Fig. \ref{fig:testcase}. It represents a meandering course for the robot, which is bounded by hockey board elements. 

Each of the board elements is 2\,m long and staggered at intervals of 2\,m. In this experiment, they only serve to delimit the course for the robot and should not have any attenuation characteristics. Wood or plastic is therefore recommended as the material. Eight \ac{STING} terminal devices are positioned in a grid with 4\,m spacing at the edge of the test field. These are mounted on tripods at the height of 1\,m. The antennas are aligned in such a way that there is a line of sight to all other terminals. The Management control station of the \ac{STING} system, as well as the operator, can be arranged arbitrarily. Since the robotic system is operated from a distance, it is recommended that the operator is protected from view, for example, by a mobile partition. In Fig. \ref{fig:testcase}, the operator is positioned at the right end of the course.
\subsection{Robot Under Test}
\begin{figure}[htb]
	\centering
	\includegraphics[width = .40\textwidth, trim = 0mm 0mm 0mm 0mm, clip]
	{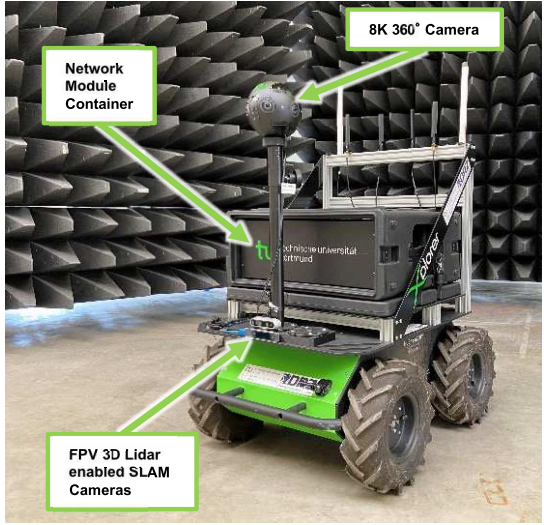}
	\caption{The Xplorer robot platform used as \ac{SUT} in the conducted test case. The robot is equipped with payloads allowing teleoperation.}
	\label{fig:xplorer}
\end{figure}

The system under test, which is visible in Fig.~\ref{fig:xplorer} is our Xplorer robot, a mid-sized \ac{UGV} based on the Clearpath Husky platform, equipped with various payloads for search and rescue missions and operated using \ac{ROS}.

The most noticeable of these payloads is the towering surround camera system, consisting of six wide-angle cameras, producing the same number of individual streams, which are then processed into one 360° perspective of the robot's surroundings. This perspective allows the rescuers to effectively survey the environment independently of the robot's current orientation.\\
A perspective more suitable for the remote control of the robot is provided by a dedicated \ac{FPV} camera mounted at the robot's front and publishes a stream of \ac{MJPEG} compressed video frames via \ac{ROS} topics.\\
Other sensors are available on the Xplorer, but only the two aforementioned were solicited in the test case in order to keep it succinct. The \ac{FPV} camera stream provides the necessary feedback for teleoperation while the traffic generated by the 360° stream allows assessing the available network datarate\\
Regarding network communications, the Xplorer is equipped with a SKATES module \cite{GuldenringSKATESInteroperableMultiConnectivity2020} which provides robust, interoperable multi-connectivity by distributing its communication links over multiple radio access technologies. In the context of this test case, however, the SKATES module was configured to rely on Wi-Fi only.
\subsection{Test Procedure}
\label{sec:procedure}
The Wi-Fi systems used for communication between the robot and the operator, as well as for the \ac{STING} system, are set to the same 20 MHz channel. A channel should be selected that is not occupied by other Wi-Fi networks not belonging to the system.
In the experiment, the robotic system is controlled several times through the course shown in Fig. \ref{fig:testcase} using teleoperation. All control and user data are transmitted. In the case of the example with the TU Dortmund robotic system, these are both the \ac{FPV} and the immersive 360° view, both of which can be used to control the system. The operator should not have a direct view of the system during the experiment.
The \ac{STING} interference communication is always switched on directly before the start of the run and then switched off again to set the configuration for the next run. The general configuration of the \ac{STING} terminal devices is summarized in Table \ref{tab:parameters}.

\begin{table}[h]
	\centering
	\begin{tabular}{|c|c|}
	\hline
	 \textbf{Parameter}   					& \textbf{Value set}                \\ \hline
	Wi-Fi Standard SUT    					&.11ac            				    \\ \hline
	Wi-Fi Standard STING  					&.11ac				               	\\ \hline 
	Active \ac{STING} end devices  			& {[}0,2,4,6,8{]}	              	\\ \hline 
	\hline
	Communication Direction 				& Uplink							\\ \hline
	Bandwidth (both systems)				& 20\,MHz							\\ \hline
  Frequency band (both systems)   & 5\,GHz              \\ \hline
  Wi-Fi Channel (both systems)          & 44                  \\ \hline
  \ac{SUT} Offered Traffic 			& 40\,Mbit/s (FPV+360$\degree$)                        \\ \hline
	Access Point (\ac{STING})					& Cisco Catalyst 9130				\\ \hline
  Access Point (\ac{SUT})					& Ubiquiti UniFi AC Mesh Pro				\\ \hline
	\ac{STING} Protocol					& UDP								\\ \hline
	\ac{STING} offered traffic	per device	& 300\,Mbit/s						\\ \hline

	\end{tabular}
	\vspace{0.1cm}
	\caption{Parameter set for case study}
	\label{tab:parameters}
\end{table}
During execution, the time required by the operator to complete the parcours is measured. The operator runs the course several times with the configurations depicted in table \ref{tab:runs}:

\begin{table}[h]
	\centering
	\begin{tabular}{|c|c|c|}
	\hline
	 \textbf{ID} & \textbf{Repetitions}   					& \textbf{Configuration}                \\ \hline
   1& 1                                & Introductory run; not tracked and no interference \\ \hline
   2& 2                                & no active interference; serves as reference for evaluation \\ \hline
   3& 2                                & 2 active \ac{STING} devices \\ \hline
   4& 2                                & 4 active \ac{STING} devices \\ \hline
   5& 2                                & 6 active \ac{STING} devices \\ \hline
   6& 2                                & 8 active \ac{STING} devices \\ \hline

	\end{tabular}
	\vspace{0.1cm}
	\caption{Performed test procedure per operator}
	\label{tab:runs}
\end{table}
\section{Test Case Evaluation}
\subsection{Influence of Interference on the robots communication link}
\label{sec:contention}
\begin{figure}[!htb]
	\centering
	\includegraphics[width=.48\textwidth]{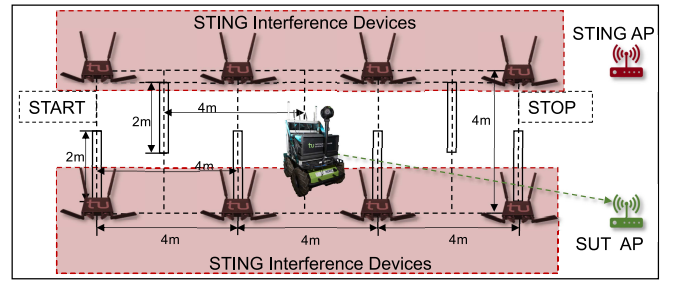}
	\caption{Constellation for interference stresstest}
	\label{fig:casestudy_interference}
\end{figure}

The following results show throughput and latency of the robot system, as well as the dropped video frames of the \ac{FPV} camera used for remote control, in a static test configuration as depicted in Fig. \ref{fig:casestudy_interference}. 

For this functional test, the \ac{SUT} remained in the middle of the scenario in order to evaluate the influence of interference on the communication of the robotic systems' camera frames. As in the test procedure described in section \ref{sec:procedure}, the number of active \ac{STING} devices was incremented in steps of two, and each configuration was active for one minute.  \\
\begin{figure}[!htb]
\begin{knitrout}
\definecolor{shadecolor}{rgb}{0.969, 0.969, 0.969}\color{fgcolor}
\includegraphics[width=0.49\textwidth]{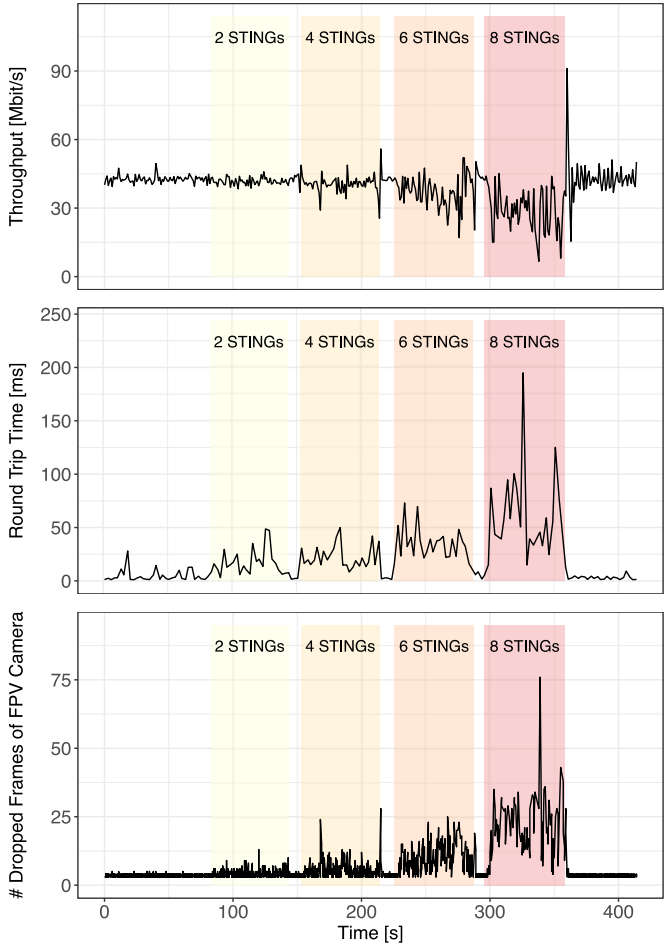} 

\end{knitrout}
\caption{Exemplary influence of STING on throughput, latency and frame drops of the \ac{FPV} Camera for different number of \ac{STING} devices}
\label{fig:contention_results}
\end{figure}

Starting with the throughput in Fig. \ref{fig:contention_results}, it can be seen that higher channel congestion with an increasing number of active \ac{STING} devices results in an overall decrease in throughput stability of the robot system. While a smaller number of active STINGs can be compensated by the camera systems catching up, six and eight active \ac{STING} devices result in a significant decrease of network throughput. 
The round trip time is also significantly increased to around 100\,ms and peaks up to 200\,ms which, together with an increased number of frame drops of the \ac{FPV} camera system, have a significant impact on the teleoperation of the robot system, as is evident in the next section.
\subsection{Influence of Interference on Remote Control Quality of Experience}
\label{sec:interference}
The test procedure described in section \ref{sec:procedure} has been conducted with four operators from our institute with different levels of experience. These result in different completion times between 43\,s and 73\,s, even without active interference. Fig. \ref{fig:contention_results} shows the completion times of every operator per run and interference constellation, as well as the distribution of completion durations of all operators per number of active \ac{STING} devices. 

\begin{figure}[!htb] 
\includegraphics[width=0.5\textwidth]{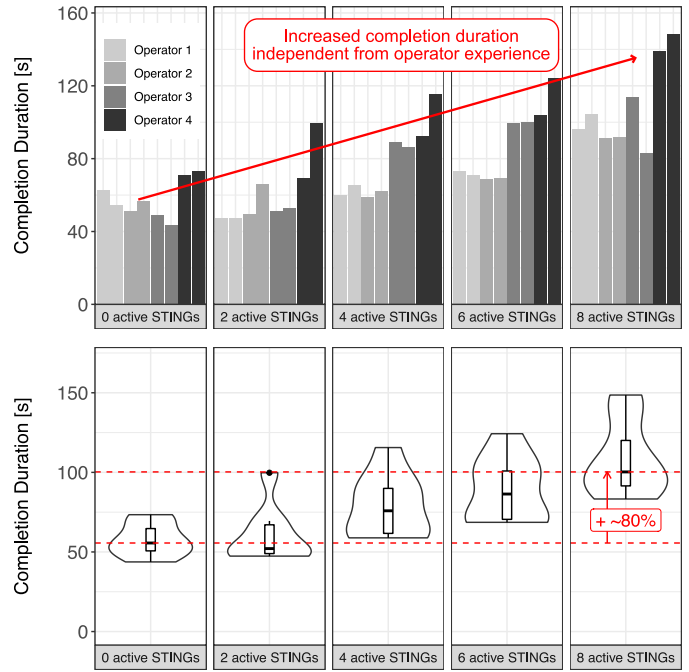}
\caption{Distribution of various operators completion time under varying STING influence}
\label{fig:latency}
\end{figure}
Results of sample executions of the test case show a strong influence of communication performance on the operability of the robot system. One run with two active \ac{STING} devices failed due to a collision with the hockey boards. From the absolute completion times of the completed runs, it is evident that all operators, while having different capabilities, need more time to finish the parcours with active interference; thus, operator experience can not fully compensate the reduced operability.
The distribution of completion times shows that a higher number of active interferers results in a significant increase of up to 80\% from 55\,s to 100\,s median completion time with eight active \ac{STING} devices compared to no interference.

The reduced completion time is mainly caused by the lack of constant camera frames, which are needed by the operator to maneuver the robot safely. With a higher interference on the wireless channel, camera frames are dropped or have a significant delay. This proves that testing the robot systems' resilience to interference is a crucial aspect when testing before real search and rescue missions. Countermeasures against interference on a channel or frequency band can be an intelligent channel switching scheme or, more reliably, incorporating an additional network interface operating in another frequency band. Such a multi-link approach is proposed and discussed in \cite{PatchouHardwareSimulationLoop2020} and \cite{GuldenringSKATESInteroperableMultiConnectivity2020}.

\section{Conclusion}
\label{sec:concl}
In this work, we propose \ac{STING}, a framework for application-aware network stress-testing. The framework is built in a modular way, enabling a variety of test methods to be applied. We showed the relevance of such a framework on a case study in the form of a rescue robotics test case. The test case consists of a parcours through which a robot has to be navigated in a teleoperation configuration without a direct line of sight of the operator. The parcours is performed multiple times with increasing interference on the used wireless channel, leading to higher latency and reduced throughput of the robot's communication link. Tests with four operators with different levels of experience showed an increased median of the completion time of about 80\%, proving the necessity of a stress-testing system for critical application scenarios. 
In future works, we want to use the system to address the coexistence of different Wi-Fi Standards, especially the impact of the current Wi-Fi 6 standard for industrial and robotic applications. 
The modular approach of the frameworks will also allow the integration of 5G modems into the \ac{STING} devices, enabling stress-testing of 5G campus networks in industrial applications.      


\section*{Acknowledgment}
Part of the work on this paper has been supported by the German Federal Ministry of Education and Research (BMBF) in the project A-DRZ (13N14857) as well as the Ministry of Economic Affairs, Innovation, Digitalization and Energy of the State of North Rhine-Westphalia (MWIDE NRW) along with the \textit{Competence Center 5G.NRW} under grant number 005-01903-0047 and \textit{5Guarantee} under grant number 005-2008-0046 and the Deutsche Forschungsgemeinschaft (DFG) within the Collaborative Research Center SFB 876 “Providing Information by Resource-Constrained Analysis”, project A4.



\bibliographystyle{IEEEtran}
%
%
%
%
%

\balance
\bibliography{2021_STING_vtc_fall_bib}

\end{document}